\documentstyle[12pt,epsf]{article}
\newcommand{\nn}{\nonumber}
\newcommand{\be}{\begin{equation}}
\newcommand{\ee}{\end{equation}}
\newcommand{\bea}{\begin{eqnarray}}
\newcommand{\eea}{\end{eqnarray}}
\newcommand{\bean}{\begin{eqnarray*}}
\newcommand{\eean}{\end{eqnarray*}}

\begin{document}
\vspace{-1cm}
\noindent
\begin{flushright}
KANAZAWA-01-02\\
KUNS-1710
\end{flushright}
\vspace{10mm}
\begin{center}
{\Large \bf 
Sfermion masses in Nelson-Strassler type of models:\\
SUSY standard models coupled with SCFTs
}
\vspace*{15mm}\\
\renewcommand{\thefootnote}{\alph{footnote}}
Tatsuo Kobayashi$^\dagger$
\footnote{E-mail: kobayash@gauge.scphys.kyoto-u.ac.jp}
and Haruhiko Terao$^{\dagger\dagger}$
\footnote{E-mail: terao@hep.s.kanazawa-u.ac.jp}
\vspace*{5mm}\\
$^\dagger$Department of Physics, Kyoto University\\ 
Kyoto 606-8502, Japan 
\vspace{2mm}\\
$^{\dagger\dagger}$Institute for Theoretical Physics, Kanazawa University\\
Kanazawa 920-1192, Japan
\end{center}
\vspace*{10mm}
\begin{abstract}
We study soft SUSY breaking parameters in the Nelson-Strassler 
type of models: SUSY standard models coupled with SCFTs.
In this type of models, soft SUSY breaking parameters including 
sfermion masses can be suppressed around the decoupling scale of 
SCFTs. We clarify the condition to derive exponential suppression
of sfermion masses within the framework of pure SCFTs.
Such behavior is favorable for degeneracy of sfermion masses.
However, the realistic sfermion masses are not quite degenerate
due to the gauge couplings and the gaugino masses in the SM sector.
We show the sfermion mass spectrum obtained in such models.
The aspect of suppression for the soft SUSY breaking parameters
is also demonstrated in an explicit model. We also give a mechanism 
generating the $\mu$-term of the Electro-Weak scale by a singlet 
field coupled with the SCFT. 
\end{abstract}
\vspace*{20mm}
\noindent

\newpage
\pagestyle{plain}
\pagenumbering{arabic}
\setcounter{footnote}{0}

\section{Introduction}

Understanding the origin of the flavor structure, {\it i.e.}
hierarchical fermion masses and mixing 
angles, is one of the most important issues in particle physics.
Actually, several types of mechanisms to realize the 
fermion mass matrices have been proposed, {\it e.g.} the 
Froggatt-Nielsen mechanism \cite{FN} and some new ideas 
concerned with extra dimensions \cite{exD,BKNY}.

Supersymmetric extension is one of 
attractive ways beyond the standard model (SM).
In supersymmetric models, realization mechanisms of 
fermion mass matrices, in general, affect the sfermion sector.
Each realization mechanism of the flavor structure 
would lead to a proper pattern of the sfermion mass matrices 
as well as supersymmetry (SUSY) breaking tri-linear couplings.
For example, within the framework of 
the Froggatt-Nielsen mechanism with gauged extra symmetries, 
sfermion masses have the so-called D-term contributions, 
which are proportional to charges of fermions under broken symmetries. 
Alternatively, if the Yukawa couplings are subject to the
infrared (IR) fixed points {\it a la} Pendelton-Ross \cite{pr}, 
we have specific relations among the soft SUSY breaking terms \cite{kt}.
Thus, study on the sfermion sector is interesting to distinguish 
several types of realization mechanisms of the flavor structure.
Furthermore, the sfermion sector has severe constraints 
due to experiments of flavor changing neutral current (FCNC) processes 
as well as CP physics \cite{fcnc}.
FCNC problems can be solved by three ways, (1) degenerate sfermion masses, 
(2) decoupling of heavy sfermion masses 
\footnote{The recent 
measurement of the muons anomalous magnetic moment \cite{mug-2}
disfavors the decoupling solution at least for the slepton sector, 
if the deviation from the prediction of the SM is indeed 
due to superpartners.} 
and (3) the alignment of the fermion and the sfermion bases.

Recently, supersymmetric standard models (SSMs) coupled with 
superconformal field theories (SCFTs) have been discussed
by Nelson and Strassler \cite{ns}.
Here the SCFT means the theory realized at a nontrivial 
IR fixed point.
Such fixed points are known to exist according to the
discussions given in ref.~\cite{seiberg}.
\footnote{See \cite{is} for a review of 
superconformal theories and their dual descriptions. 
The IR fixed points have been discussed also in \cite{BaZa,oehme1,Kubo}.}
Within this framework, quark and lepton fields 
coupled with the superconformal (SC) sector have enhanced anomalous 
dimensions due to strong gauge and Yukawa couplings 
in the SC sector around IR fixed points.
The anomalous dimensions lead to the hierarchically suppressed 
Yukawa couplings at low energy in the SSM sector even if those are 
of $O(1)$ at high energy.
Thus, this can provide with one type of 
mechanisms to generate realistic quark and lepton mass 
matrices.

Superconformal IR fixed points have more intriguing aspects 
for renormalization group (RG) behavior of SUSY breaking 
parameters. 
For example, IR behavior of softly
broken supersymmetric QCD has been studied in ref.~\cite{kkkz} and 
it has been shown that the gaugino mass and the
squark masses are exponentially suppressed around 
the IR fixed point \footnote{See also \cite{lr}.}.
Furthermore, its dual theory is described in terms of  
dual quarks and singlet (meson) fields \cite{seiberg}.
In the dual side, SUSY breaking tri-linear couplings are suppressed.
Moreover the soft scalar masses of the singlet fields
as well as  the sum of (mass)$^2$ for the dual squark and its
conjugate are found to be suppressed.

In ref.~\cite{ns}, it has been mentioned that 
the above behavior of suppressed soft scalar masses 
around the IR fixed point can be useful to avoid 
the dangerous FCNC processes.
Because it is expected that sfermion masses at least for the first and 
the second families would be quite suppressed at the decoupling scale of 
the SC sector
and that after decoupling the masses receive
radiative corrections due to gaugino masses of the SSM sector,
which are flavor-blind.
Thus, sfermion masses could be degenerate at low energy 
for any initial condition at high energy.

{}From the above phenomenological viewpoint, 
it is quite interesting to study the IR behavior of softly broken SCFTs 
and SSMs coupled to the SC sector,
and to clarify the conditions leading to exponentially suppressed SUSY
breaking terms.
That would provide with useful constraints for model building.

In this paper we study first the IR behavior of general softly broken SCFTs
by means of the so-called exact beta functions.
The conditions to realize suppressed SUSY breaking terms will be shown.
It is also shown that some fields become tachyonic in generic cases.
These aspects around the IR fixed point are useful also for phenomenology.
Next we discuss phenomenological aspects of SSMs coupled with the SC sector.
Taking account of the effects due to gaugino masses of SSMs, the sfermion
masses are found to converge to flavor dependent values.
We study this flavor dependence of the sfermion masses remained after 
suppression and show how much degeneracy between sfermion masses is
achieved finally at the weak scale.
In practice the range of scale where the SSMs couple with the SC sector
must be finite so as to generate the small but non-vanishing Yukawa 
couplings. Therefore the sfermion masses do not totally converge at the
decoupling scale. We also discuss the amount of convergence by 
demonstrating the RG flows for explicit models.
It is mentioned also  that the SCFT may resolve the $\mu$-problem in SSMs
in a natural way.
We will show a model in which soft SUSY breaking mass of a singlet coupled with 
SCFTs becomes tachyonic and appears of the weak scale automatically.
If this singlet couples to the Higgs fields, then the $\mu$-term may be
generated through vacuum expectation value of the weak scale.

This paper is organized as follows.
In section 2, we study IR behavior of pure SCFTs 
with soft SUSY breaking terms and show the condition for 
suppressed sfermion masses.
We also give speculative considerations on SCFTs with nonrenormalizable 
couplings and corresponding SUSY breaking terms.
In section 3 we study 
SSMs coupled with the SC sector.
In subsection 3.1 we give a brief review on the set-up of
Nelson-Strassler models, and also a constraint for the decoupling
scale of the SC sector is given.
In subsection 3.2 it is found that 
gaugino masses in the SSM sector have important meaning for
exponential suppression of 
soft scalar masses, and that is significant also from the viewpoint 
of FCNC constraints.
Within the framework of minimal SSM (MSSM), we give 
numerically how strongly sfermion masses are degenerate at the weak scale.
Also a typical mass spectrum is shown in subsection 3.3.
In section 4 we consider explicit models showing the desired suppression.
After discussing the types for such models in subsection 4.1,
we demonstrate typical RG-flows in an illustrating model in subsection 4.2.
The convergence of the sfermion masses are examined also there.
In section 5 the $\mu$-problem is discussed. 
Section 6 is devoted to conclusion and discussions.

\section{
Exact results for soft masses in SCFT}

\subsection{Beta functions}
In this section we are going to  discuss IR behavior of the soft
parameters added to generic SCFTs.
In particular soft scalar masses will be found to satisfy interesting
sum rules.
Our argument is based on the explicit form of the beta functions
for soft parameters \cite{yamada}-\cite{softbeta}.
Therefore we first review the exact beta functions of
general softly broken supersymmetric gauge theories in this 
subsection.

Let us begin with the gauge coupling and the corresponding
gaugino mass.
The holomorphic gauge coupling $S=1/2g^2_h$ satisfies the RG equation (RGE):
\be
\mu\frac{dS}{d\mu} = \frac{1}{16\pi^2}(3 T_{\rm G} - \sum_i T_i),
\ee
where $T_i$ is the Dynkin index and $T_G$ denotes 
the Dynkin index (quadratic Casimir) of adjoint representation.
The physical coupling $g$ is related to the holomorphic coupling
through the general formula:
\be
8\pi^2(S+S^{\dagger}) - \sum_iT_i\ln Z_i =
\frac{1}{\alpha} + T_{\rm G}\ln \alpha + 
\sum_{n>0}a_n \alpha^n \equiv F(\alpha),
\ee
where $\alpha=g^2/8\pi^2$ and $Z_i$ denotes the wave function 
renormalization of the chiral superfield $\phi^i$.
The coefficients 
$a_n$ are the scheme-dependent constants and the NSVZ scheme \cite{novikov} 
is given by
$a_n=0$.
{}From this relation the beta function for $\alpha$ is given exactly as
\be
\beta_{\alpha}=\mu\frac{d\alpha}{a\mu}
=\frac{1}{F'(\alpha)}[3T_{\rm G} - \sum_iT_i(1 - \gamma_i)],
\label{eq:gaugebeta}
\ee
where the anomalous dimension $\gamma_i$ is defined by
\be
\gamma_i = - \frac{d \ln Z_i}{d \ln \mu}.
\ee
Here we assume the wave function renormalization to be
diagonal just for simplicity.

The gaugino mass can be incorporated with the gauge coupling by
superfield extension. The holomorphic coupling is extended as
\be
\tilde{S} = \frac{1}{2g^2_h}(1-M_h \theta^2).
\ee
Here $\tilde{S}$ satisfies the same RGE as $S$ does.
On the other hand we extend also the wave function renormalization factors
$Z_i$ and the physical gauge coupling to real superfields as
\bea
\tilde{Z}_i &=&
\tilde{Z}_{\phi i}(1-m_i^2\theta^2\bar{\theta}^2)
\tilde{\bar{Z}}_{\phi  i}, 
\nn \\
\tilde{\alpha} &=&
\alpha\left(1 + M\theta^2 + \bar{M}\bar{\theta}^2 +
(2M\bar{M} + \Delta_g) \theta^2\bar{\theta}^2\right),
\label{eq:spurion}
\eea
where $M$ and $m_i^2$ give the gaugino mass and the soft scalar
masses respectively. 
We have extracted the chiral and the anti-chiral parts of $\tilde{Z}_i$ as
$\bar{Z}_{\phi i}$ and $\tilde{\bar{Z}}_{\phi i}$ for the wave function
renormalization of the (anti-)chiral matter fields.
Here $\Delta_g$ is determined by consistency with the extended 
relation
\be
8\pi^2(\tilde{S}+\tilde{S}^{\dagger}) - \sum_iT_i\ln \tilde{Z}_i =
F(\tilde{\alpha}),
\ee 
and is found out to be
\be
\Delta_g = \frac{1}{\alpha F'(\alpha)}
\left[ \sum_iT_im_i^2 - (\alpha^2 F'(\alpha))'M \bar{M} \right].
\ee
In the NSVZ scheme $\Delta_g$ is given by \cite{kkz, jjp, kv}
\be
\Delta_g = - \frac{\alpha}{1-T_{\rm G} \alpha}
\left[\sum_iT_im_i^2 - T_{\rm G}M \bar{M}
\right].
\label{Delta_g}
\ee

The beta function of the gaugino mass can be derived from the
extended relation by expanding with $\theta^2$ and found to be
\be
\mu\frac{dM}{d\mu} 
=
\left. 
\mu\frac{d \ln\tilde{\alpha}}{d\mu}
\right|_{\theta^2}
= 
\left.
\frac{1}{\tilde{\alpha}F'(\tilde{\alpha})}
[3T_{\rm G} - \sum_iT_i(1 - \tilde{\gamma}_i)]
\right|_{\theta^2},
\ee
where the extended anomalous dimension $\tilde{\gamma}_i$ is
given by 
\be
\tilde{\gamma}_i = - \frac{d \ln \tilde{Z}_i}{d \ln \mu}
\equiv \gamma_i + \gamma^{(1)}_i \theta^2 + \bar{\gamma}^{(1)}_i \bar{\theta}^2
+ \gamma^{(2)}_i \theta^2 \bar{\theta}^2.
\ee

Next let us consider the Yukawa couplings and the tri-linear couplings
given by the superpotential
\be
W= \frac{1}{6}(y^{ijk} - h^{ijk}\theta^2 )
\phi^i \phi^j \phi^k.
\ee
The SUSY breaking tri-linear coupling $h^{ijk}$ is often written 
as $h^{ijk} = y^{ijk}A^{ijk}$, where $A^{ijk}$ are called as A-terms.
Because of non-renormalization of the superpotential,
the holomorphic couplings $Y^{ijk}=y^{ijk} - h^{ijk}\theta^2 $
are renormalized by the chiral superfield $\tilde{Z}_{\phi i}$
as
\be
Y^{ijk}_{\rm bare} =
Y^{ijk} \tilde{Z}^{-1}_{\phi i} \tilde{Z}^{-1}_{\phi j} 
\tilde{Z}^{-1}_{\phi k}.
\ee
By noting that the chiral superfields are represented as
\be
\tilde{Z}^{-1}_{\phi i}=
Z_i^{1/2} + \left. \tilde{Z}_i \right|_{\theta^2} \theta^2,
\ee
we can immediately derive the beta functions
for the Yukawa couplings and the tri-linear couplings
as
\bea
\beta_{y}^{ijk}&=&\mu\frac{dy^{ijk}}{d\mu}
= \frac{1}{2}(\gamma_i + \gamma_j + \gamma_k)y^{ijk},  \nn \\
\beta_{h}^{ijk}&=&\mu\frac{dh^{ijk}}{d\mu}
= \frac{1}{2}(\gamma_i + \gamma_j + \gamma_k)h^{ijk}
-(\gamma^{(1)}_i + \gamma^{(1)}_j + \gamma^{(1)}_k)y^{ijk}.
\eea

It has been known that the wave function superfields $\tilde{Z}_i$
are also given by the extension of 
$Z_i(\alpha, y^{ijk}, \bar{y}_{ijk})$ \cite{kazakov}:
\be
\tilde{Z}_i = Z_i(\tilde{\alpha}, \tilde{y}^{ijk}, \tilde{\bar{y}}_{ijk}),
\ee
where the extended Yukawa couplings $\tilde{y}^{ijk}$ are defined by
\be
\tilde{y}^{ijk} = Y^{ijk}
+ \frac{1}{2}(m_i^2 +m_j^2 +m_k^2)y^{ijk} \theta^2\bar{\theta}^2.
\ee
Therefore the superfields $\tilde{Z}_i$ are given explicitly
in terms of the rigid factor $Z_i$ as
\be
\tilde{Z}_i = Z_i + D_1 Z_i \theta^2 + \bar{D}_1 Z_i \bar{\theta}^2 
+ D_2 Z_i \theta^2 \bar{\theta}^2,
\ee
where $D_1$ and $D_2$ are the differential operators defined by
\bea
D_1&=& M \alpha\frac{\partial}{\partial \alpha}
- h^{ijk}\frac{\partial}{\partial y^{ijk}}. \nn \\
D_2&=& \bar{D}_1 D_1 + 
(M\bar{M} + \Delta_g)\alpha\frac{\partial}{\partial \alpha}
+\frac{1}{2}(m_i^2 +m_j^2 +m_k^2)
\left(
y^{ijk}\frac{\partial}{\partial y^{ijk}} +
\bar{y}_{ijk}\frac{\partial}{\partial \bar{y}_{ijk}}
\right).
\eea
Here it will be helpful for the later discussions to note that 
$\tilde{\alpha}_{y}^{ijk} = |\tilde{y}^{ijk}|^2/8\pi^2$ 
satisfies the same form of renormalization as the rigid one:
\be
\tilde{\alpha}_{y {\rm bare}}^{ijk} = \tilde{\alpha}_{y}^{ijk}
\tilde{Z}^{-1}_i\tilde{Z}^{-1}_j\tilde{Z}^{-1}_k.
\ee
Moreover, we find the beta functions for the scalar masses
also by the superfield extension
as
\bea
\beta_{m^2_i} \equiv
\mu\frac{dm_i^2}{d\mu} &=&
-\left. 
\mu\frac{d \ln\tilde{Z}_i}{d\mu} \right|_{\theta^2\bar{\theta}^2} \nn \\
&=& \gamma^{(2)}_i \nn \\
&=& D_2 \gamma_i.
\label{beta-soft-m}
\eea

\subsection{RG flows around the IR stable fixed points}
First let us consider the IR fixed point of the rigid beta functions
where the SCFT realizes.
The beta functions for the gauge coupling and the Yukawa couplings
vanish, {\it i.e.}
$\beta_\alpha = \beta^{ijk}_y =0$, when the anomalous dimensions
satisfy the following conditions:
\bea
& &\sum_i T_i \gamma_i = 3T_G -\sum_i T_i, \nn \\
& &\gamma_i +\gamma_j + \gamma_k =0,
\label{eq:fpcond}
\eea
for each Yukawa coupling.
We may wonder that these conditions are insufficient to determine the
fixed points, since the Yukawa couplings are complex in general. 
However the phase of the Yukawa coupling is not renormalized by the
real wave function renormalization.
Also the anomalous dimensions are actually independent of the phases,
since they are found to satisfy \cite{jj}
\be
y^{ijk} \frac{\partial \gamma}{\partial y^{ijk}}
= \bar{y}_{ijk} \frac{\partial \gamma}{\partial \bar{y}_{ijk}}.
\ee
As a result the phases of the Yukawa couplings are completely undetermined 
in all order of perturbation.
This is similar to the behavior of $\theta$-parameter 
in generic gauge theories.
On the other hand, however, the (ir)relevance of the couplings
concerns only with 
evolution of their absolute values.
Therefore we should rather consider the real couplings
$\alpha_{y}^{ijk} = |y^{ijk}|^2/8\pi^2$.

Now we assume existence of the IR attractive non-trivial 
fixed points $(\alpha_*, \alpha^{ijk}_{y*})$.\footnote{ 
We do not consider the possibility of fixed lines.}
Then generic low energy effective theories turn out to be SCFTs 
subject to these fixed points.
Around the IR attractive fixed points, both of the gauge coupling and
the Yukawa couplings should be irrelevant.
If we take infinitesimal variations from the fixed point:
$\alpha=\alpha_* + \delta \alpha, \alpha^{ijk}_{y}=\alpha^{ijk}_{y*}+ 
\delta\alpha^{ijk}_{y}$, 
then the variations are subject to the linear differential 
equations
\bea
\mu \frac{d\delta\alpha}{d\mu} &=&
\left(\frac{\partial \beta_{\alpha}}{\partial \alpha}\right)_*
\delta \alpha+
\left(\frac{\partial \beta_{\alpha}}{\partial \alpha^{ijk}_{y}}\right)_*
\delta \alpha^{ijk}_{y}, \nn \\
\mu \frac{d\delta\alpha^{lmn}_{y}}{d\mu} &=&
\left(\frac{\partial \beta^{lmn}_{\alpha_{y}}}{\partial \alpha}\right)_*
\delta \alpha+
\left(\frac{\partial \beta^{lmn}_{\alpha_{y}}}
{\partial \alpha^{ijk}_{y}}\right)_*
\delta \alpha^{ijk}_{y},
\label{d-a-ay}
\eea
where the asterisk represents evaluation at the fixed point.
The irrelevance of these couplings means that the eigenvalues of these 
equations are all positive.

Next let us consider the IR behavior of the gaugino mass and the tri-linear
couplings \cite{kkkz,lr}.
As we have already seen, the beta functions for these 
couplings can be obtained by the Grasmannian expansion.
Specially the extended couplings $\tilde{\alpha}$ and
$\tilde{\alpha}^{ijk}_{y}$ satisfy the same form of the
RG equations:
\bea
\mu \frac{d\tilde{\alpha}}{d\mu} &=&
\beta_{\alpha}(\tilde{\alpha}, \tilde{\alpha}_y), \nn \\
\mu \frac{d\tilde{\alpha}^{ijk}_{y}}{d\mu} &=&
\beta^{ijk}_{\alpha_y}(\tilde{\alpha}, \tilde{\alpha}_y).
\eea
As a result, $\alpha_* M$ and $-\bar{y}_{ijk*}h^{ijk}/8\pi^2$, 
as well as their complex conjugates, 
are found to satisfy the same linear differential equations
for $\delta\alpha$ and $\delta \alpha^{ijk}_{y}$ given by 
(\ref{d-a-ay}) around the fixed point.
Therefore both of $M$ and $h^{ijk}$ acquire negative anomalous 
dimensions and decrease exponentially towards the IR region.
Note that this does not always mean that these couplings are irrelevant
in Wilson's sense, since they are dimensionful.

We also regard the $\theta^2\bar{\theta}^2$ components
of the extended couplings as the infinitesimal variations.
Since $M$ and $h^{ijk}$ vanish at the IR regime, the variations
given by
\bea
\delta\alpha &=& 
\frac{1}{F'(\alpha_*)} \sum_i T_i m_i^2 \theta^2\bar{\theta}^2 ,\nn \\
\delta\alpha^{ijk}_y &=& 
\alpha^{ijk}_{y*}
(m_i^2+m_j^2+m_k^2) \theta^2\bar{\theta}^2,
\eea
satisfy eq.~(\ref{d-a-ay}).
This shows that $\sum_i T_i m_i^2$ as well as $m_i^2+m_j^2+m_k^2$ 
corresponding to the Yukawa couplings $y^{ijk}$ decrease exponentially 
towards the IR regime.
By using the IR behavior of the soft parameters clarified so far,
it is seen that the beta functions for soft scalar masses also decrease 
exponentially.
Consequently we find that the soft scalar masses approach to the
constant values satisfying 
\bea
& &\sum_i T_i m_i^2=0,  \nn \\
& &m_i^2+m_j^2+m_k^2=0, 
\label{eq:sumrule}
\eea
for each Yukawa coupling. 
Each IR value $m_i$ is heavily dependent on the initial soft parameters.
However these relations among them must be universal.
In the case that the anomalous dimensions $\gamma_i$ are
completely determined by eq.~(\ref{eq:fpcond}),
the above equations (\ref{eq:sumrule}) lead to the vanishing
IR soft scalar masses for the corresponding masses.
This happens whenever the anomalous dimensions of the fields
can be uniquely determined from an R-symmetry, since the dimension
of the field must be given by the R-charge in SCFT \cite{ns}.

We mention the dual SQCD as a special case. 
The theory contains the magnetic quark pairs $(q, \bar{q})$ and 
a gauge singlet $M$ and 
the Yukawa coupling of them is unique, $W=yq\bar q M$.
Therefore the soft masses of them should behave as
\footnote{Similar discussions of suppressing the sfermion masses 
in the dual side have been done also in ref.~\cite{kkkz}.
} 
\be
(m_q^2, m_{\bar{q}}^2, m^2_{M}) 
\stackrel{\mu \rightarrow 0}{\longrightarrow}
m^2(1, -1, 0).
\ee
If we assume $m_q^2=m_{\bar{q}}^2$ as the initial condition, then
all scalar masses are exponentially suppressed.

\subsection{Higher dimensional interactions}
The higher dimensional operators can be turned into ones
relevant to the large anomalous dimensions at the fixed point.
Therefore we should include such operators as well to find
the IR stable fixed points in general.
However we cannot apply the RG framework for the renormalizable
theories discussed so far.
If there were found the Wilson RG respecting the gauge symmetries 
and supersymmetry simultaneously, it would give a suitable 
framework instead.
Here we naively assume such a framework and discuss the IR behavior
of the soft parameters around such fixed point somehow
speculatively.

Suppose that the superpotential of SCFT contains also higher
dimensional operators such as
\be
W = \sum \frac{1}{n!} 
\frac{y^{i_1 i_2\cdots i_n}}{\mu^{n-3}} \phi^{i_1}\phi^{i_2}\cdots \phi^{i_n}.
\label{eq:generalint}
\ee
The non-renormalization for the superpotential and the gauge coupling 
may well be supposed to remain intact.
\footnote{
Perturbative nonrenormalization theorem applied to the nonrenormalizable
theories has been presented in ref. \cite{weinberg}.
On the other hand non-renormalization is not maintained non-perturbatively
in general, {\it e.g.} the Affleck-Dine-Seiberg superpotential.
We assume that such corrections are absent in the following argument.
}
Then we write the Wilsonian effective Lagrangian as
\be
{\cal L} = \int d^4\theta K(\phi^i, \phi^{i \dagger}, V)
+ \int d^2\theta \frac{1}{16g_h^2}\mbox{\rm tr}W^{\alpha}W_{\alpha}
+ \int d^2\theta W(\phi^i) + h.c.,
\ee
where the superpotential $W$ is given by eq.~(\ref{eq:generalint}).
The K\"{a}hler potential $K$ given generally as
\be
K(\phi^i, \phi^{i \dagger}, V) = 
Z_i \phi^{i \dagger} e^{-V} \phi^i + \sum \kappa_n {\cal O}_n,
\ee
contains generic operators ${\cal O}_n$ allowed by symmetries.
It should be noted that the wave function renormalization factors $Z_i$ 
depend also on the effective couplings $\kappa_n$ as well as
other couplings in the Wilson RG.

The gauge beta function is given in the same way as eq.~(\ref{eq:gaugebeta})
except the fact that the anomalous dimension $\gamma_i$ is defined
from the generalized wave function renormalization $Z_i$.
The beta functions for the  couplings $y^{i_1 i_2 \cdots i_n}$ 
are also given by
\be
\beta_{y}^{i_1 i_2 \cdots i_n}=\mu\frac{dy^{i_1 i_2 \cdots i_n}}{d\mu}
= (n-3)y^{i_1 i_2 \cdots i_n}
+\frac{1}{2}(\gamma_{i_1} + \gamma_{i_2} + \cdots + \gamma_{i_n})
y^{i_1 i_2 \cdots i_n}.
\ee
The beta functions for $\kappa_n$ are unknown though.
All these beta functions are required to vanish at the fixed points.
If the fixed point action contains the higher dimensional interaction
$y^{i_1 i_2\cdots i_n} \phi^{i_1}\phi^{i_2}\cdots \phi^{i_n}$, then
we impose 
\bea
& &\sum_i T_i \gamma_i^* = 3T_G - \sum_i T_i, \nn \\
& &\gamma_{i_1}^* + \gamma_{i_2}^* + \cdots + \gamma_{i_n}^* = -2(n-3),
\eea
as the necessary condition. 
If the fixed point is IR attractive, then all the eigenvalues of the
linearized beta functions for the infinitesimal variation from the
fixed point values must be positive.

Now we shall consider to incorporate the SUSY breaking parameters by
applying the spurion method.
We introduce the chiral superfield
$Y^{i_1 i_2\cdots i_n}=y^{i_1 i_2\cdots i_n}-h^{i_1 i_2\cdots i_n}\theta^2$
and the real superfields $\tilde{\kappa}_n$ adding to $\tilde{Z}_i$
and $\tilde{\alpha}$ defined by eq.~(\ref{eq:spurion}).
Here suppose the wave function superfield $\tilde{Z}_i$ is simply given 
by the extension as
\be
\tilde{Z}_i = Z_i(\tilde{\alpha}, \tilde{y}^{ij\cdots k},
\tilde{\bar{y}}_{ij\cdots k}, \tilde{\kappa}_n),
\ee
where the extended couplings $\tilde{y}^{ij\cdots k}$ are
defined by
\be
\tilde{y}^{ij\cdots k}=Y^{ij\cdots k} + 
\frac{1}{2}(m_i^2 + m_j^2 + \cdots + m_k^2)y^{ij\cdots k} 
\theta^2\bar{\theta}^2.
\label{extyukawa}
\ee
The reasoning of this extension is the same for the Yukawa coupling.
Then the beta functions for 
$|\tilde{y}^{ij\cdots k}|^2$ as well as $\tilde{\alpha}$ can be
given by extending the couplings in the rigid beta functions
for $|y^{ij\cdots k}|^2$.
Since the fixed point is IR attractive, the $\theta^2 \bar{\theta}^2$ term
in the extended couplings given by (\ref{extyukawa}) decreases exponentially
again.
Namely we could obtain the extended sum rule at IR as
\be
m_i^2 + m_j^2 + \cdots + m_k^2 \rightarrow 0.
\ee

\section{SSMs coupled with SCFT}

\subsection{Yukawa hierarchy}

Here we give a brief review on the mechanism to 
realize hierarchically suppressed Yukawa couplings 
following ref.~\cite{ns}.
We assume two sectors: One is the SSM sector, which has  
the gauge group $G_{SM} = SU(3) \times SU(2) \times U(1)_Y$ or 
an extended group, and three families of quarks and leptons 
as well as Higgs fields $H_{u,d}$.
The $i$-th family of them are denoted by $q_i$ representatively, 
and they have ordinary Yukawa couplings $y^{ij}_{u,d}q^i_{L} q^j_{R} H_{u,d}$.
The other sector is the SC sector, which has the gauge group $G_{SC}$
and matter fields, which are denoted by $\Phi^r$ representatively.
The SC-sector matter fields also have their couplings 
$\lambda'^{r_1 r_2\cdots r_n}\Phi^{r_1} \Phi^{r_2} \cdots \Phi^{r_n}$, 
and 
the first two families of $q^i$ are assumed to have Yukawa couplings 
with $\Phi^r$, i.e. $ \lambda^{rsi} \Phi^r \Phi^s q^i$.
In the small $\tan \beta$ scenario, the bottom quark and 
tau lepton as well as the down sector Higgs field $H_d$ 
must be coupled to $\Phi^r$.
Altogether we have the following superpotential,
\begin{equation}
W = y^{ij}_{u,d}q^i_{L} q^j_{R} H_{u,d} + 
\lambda'^{r_1 r_2\cdots r_n}\Phi^{r_1} \Phi^{r_2} \cdots \Phi^{r_n} + 
\lambda^{rsi}\Phi^r \Phi^s q^i.
\end{equation}
The SSM matter fields $q^i$ and $H_{u,d}$ are assumed to be 
singlets under $G_{SC}$.
Hence, some of SC matter fields $\Phi^r$ must have 
nontrivial representations under $G_{SM}$ to 
allow Yukawa couplings $ \lambda^{rsi} \Phi^r \Phi^s q_i$.
The gauge couplings of the SSM sector and SC sector are 
denoted by $g_a$ ($a=1,2,3$) and $g'$, respectively, and the gauge 
group $G_{SC}$ is assumed to be strongly coupled.
On top of that, as said in the previous section, 
it is expected that the SC sector has a non-trivial IR fixed point.
Here we assume that the gauge couplings $g_a$ of the SSM sector are weak 
compared with $g'$.
Then we neglect $g_a$ and Yukawa couplings $y^{ij}_{u,d}$ of the SSM sector 
for calculations of the fixed point for $g'$, $\lambda^{rsi}$ and 
$\lambda'^{r_1 r_2\cdots r_n}$, that is, 
$\beta_{\alpha'}=\beta_\lambda = \beta_{\lambda'} = 0$, where 
\begin{eqnarray}
\beta_{\alpha'} &=& {1 \over F(\alpha') }
[3 T_G -\sum_r T_r(1 - \gamma_r)] \nonumber \\
\beta_\lambda^{rsi} &= &\lambda^{rsi}(\gamma_r + \gamma_s + \gamma_i) 
\label{fp2:cond} \\
\beta_{\lambda'}^{r_1 r_2\cdots r_n} &= &
\lambda^{r_1 r_2\cdots r_n}(\gamma_{r_1} + \gamma_{r_2} + \cdots 
+ \gamma_{r_n}). \nonumber
\end{eqnarray}
Through this procedure, the anomalous dimensions $\gamma_i$ of the SSM matter 
fields $q^i$ are obtained by 
fixed point values of $g'$ and $\lambda^{rsi}$ and 
$\lambda'^{r_1 r_2\cdots r_n}$, 
and in general, large.
In particular, the anomalous dimension $\gamma_i$ is fixed 
to be a definite value in the certain case as discussed 
in the previous section, and also in that case 
the corresponding sfermion mass is exponentially suppressed.
Thus, we have the following beta-function of $y^{ij}_{u,d}$, 
\begin{equation}
\beta^{ij}_{y_{u,d}} = {1 \over 2}y^{ij}_{u,d} 
(\gamma_{Li}+\gamma_{Rj}+\gamma_{H_{u,d}}),
\end{equation}
and the Yukawa coupling $y^{ij}$ at the decoupling energy scale $M_c$ 
of the SC-sector is obtained 
\begin{equation}
y^{ij}_{u,d}(M_c) = y^{ij}_{u,d}(M_0) \left( {M_c \over M_0 }\right)^{(
\gamma_{Li}+\gamma_{Rj}+\gamma_{H_{u,d}})/2},
\end{equation}
where $y^{ij}_{u,d}(M_0)$ is an initial condition at $M_0$.
The factor $M_c/M_0$ gives the suppression factor.
Thus, even if $y^{ij}(M_0) = O(1)$ for most of $(i,j)$, 
we can have hierarchical Yukawa matrices by powers of 
large anomalous dimensions $\gamma_i$.\footnote{This form is 
similar to Yukawa couplings with power-law behavior due to 
Kaluza-Klein modes in extra dimensions, where extra dimensions 
actually plays a role similar to anomalous dimensions \cite{BKNY} 
and in this case FCNC problems could  
be solved by the IR alignment mechanism \cite{ky}.}
Note that $y^{ij}(M_C)$ itself is not a fixed point or 
its value is not fixed, but its suppression factor is fixed.
Here large anomalous dimensions play a  role similar to  
$U(1)$ charges of the Froggatt-Nielsen mechanism with 
an extra $U(1)$ symmetry.
Resultant Yukawa matrices have the same form as the 
Froggatt-Nielsen mechanism.
To obtain realistic Yukawa matrices, we need non-degenerate 
anomalous dimensions, $\gamma_i \neq \gamma_j$.

The decoupling energy scale $M_c$ is obtained by 
mass terms of $\Phi^r$.
In general, families can have different decoupling energy scales 
each other, because they couple with different fields $\Phi^r$.
However, here we restrict ourselves to the universal 
decoupling scale $M_c$ for simplicity.
The discussions in the following sections can be 
extended to the case with non-universal decoupling scales.
Such mass terms for decoupling can be generated by another 
dynamical mechanism.

The decoupling energy scale $M_c$ should not be as low as 
the weak scale.
One constraint for $M_c$ comes from the fact that 
$\Phi^r$ are charged under $G_{SM}$ and inclusion of 
such extra matter fields change beta-function coefficients of 
$G_{SM}$ to be asymptotically non-free.
In that case, the gauge couplings would be strong at 
a high energy scale and comparable with $g'$ of the SC sector.
Then, the above fixed point calculations with neglecting $g_a$ are 
not reliable and the above mechanism to produce 
hierarchically suppression Yukawa couplings would be spoiled.
For example, here we assume that the gauge couplings of $G_{SM}$ 
should not blow up below the GUT scale $M_X = 2 \times 10^{16}$ GeV.
Then we take the case that the beta-function coefficient of 
$SU(3)$ is obtained by $b_3 = -3$ (just like the MSSM) below
$M_c$ up to $M_Z$ and above $M_c$ extra matter fields $\Phi_a$ 
contribute to it 
as $b_3 = -3 +x$.
Fig.~1 shows the curve of ($M_c,x)$ corresponding to 
the gauge coupling $g_3$, which blows up at $M_X$.
We have used the one-loop beta-function.
The region above the curve corresponds to the region where 
$g_3$ blows up below $M_X$.

\begin{center}
\input Mc-b.tex

Figure 1: Blow-up of $g_3$
\end{center}

Also the gauge coupling unification is spoiled if 
we add generic extra matter fields.
However, the coupling unification still holds at $M_X$ in 
the case that we add extra matter such that 
the beta-function coefficients shift universally from 
the values of MSSM, $b^{MSSM}_a \rightarrow b^{MSSM}_a + x$ 
at $M_c$.
We assume this situation in the whole of this section.
A value of the unified coupling $\alpha_X$ changes from the 
value for the MSSM and in general, it becomes strong.

In the previous section, we have seen that soft scalar masses 
are exponentially suppressed around the IR fixed point 
in the case that the corresponding anomalous dimension $\gamma_i$  
is determined definitely by eq.~(\ref{fp2:cond}).
That is favorable for FCNC constraints.
Because after such suppressions at $M_c$, we have radiative corrections 
due to the gaugino masses of the SSM sector, which are 
flavor-blind.
Actually, such possibility has been mentioned 
in ref.~\cite{ns}.
However, in the previous section we have considered 
the pure SC sector.
It is important to study effects of finite gauge couplings 
and gaugino masses of the SSM sector for 
realistic models.
That is the purpose of the following section.
Actually, we shall show the gauge couplings and the gaugino masses 
of the SSM sector play an important role.

\subsection{Degeneracy of sfermion masses}

In section 2, we have shown that within the framework of 
pure SCFTs soft scalar masses as well as 
gaugino masses and $A$-parameters decrease exponentially at $M_c$ 
in the case that the corresponding anomalous dimensions are determined 
definitely.
That is favorable from the viewpoint of FCNC problems.
Because that would provide degenerate sfermion masses 
at the weak scale by flavor-blind radiative corrections due to 
gaugino masses of the SM sector.
However, in a realistic case we have to examine two points 
for SSMs coupled with SCFTs: 
One is that we have to take into account effects due to 
gauge couplings and gaugino masses of the SM sector.
The other point is that a running region is finite.
The former point is considered in this subsection, while 
in subsection 4.2 the latter shall be discussed by use of an 
illustrating model.

For concreteness, we consider the case that 
$M_c$ is less than $M_X$ and below $M_C$ we have 
the same matter content as the MSSM.
It is possible to assume $M_C > M_X$, that is, 
the Nelson-Strassler mechanism works above $M_X$.
It is easy to extend the following calculations to such cases, 
although results are GUT model-dependent.

Here we denote gaugino masses of the SM sector as $M_a$.
We assume the universal gaugino mass $M_a = M_{1/2}$ at 
the GUT scale $M_X$.
Recall that the gauge coupling is unified at $M_X$ 
in the case with the beta-function coefficients 
$(b_1,b_2,b_3)=(b^{MSSM}_1+x,b^{MSSM}_2+x,b^{MSSM}_3+x)$ 
and we are taking such case.
It holds that $M_a/\alpha_a$ is a RG invariant.
Suppose that the theory is regarded as SCFT at the scale of
$M_c < \mu < M_X$.
In the RG equations of soft scalar masses we ignore the gaugino mass
and A-parameters $A_\lambda$ of the SC sector
because they decrease rapidly. 
We shall be back to this point later.
Then the RG equations for soft scalar masses are written down as
\bea
\mu \frac{d m^2_{i}}{d\mu} &=& 
{\cal M}_{ij}m^2_{j} - C_{ia} \alpha_a M^2_a \nonumber \\
&=&{\cal M}_{ij}\left(
m^2_{j} - {\cal M}^{-1}_{jk} C_{k a}\alpha_a M^2_a
\right),
\label{rg-m-M}
\eea
where $C_{ia}$ is a quadratic Casimir.
In the pure SCFT limit $\alpha_a \rightarrow 0$, 
the second term vanishes and soft scalar masses 
continue to decrease exponentially.
However, the exponentially suppressing behavior is stopped 
by the second term.
Evolution of $\alpha_a M^2_a$ is small compared with 
exponential running of the soft scalar masses.
Thus, the term $\alpha_a M^2_a$ could be treated constant 
during the exponential running of the soft scalar masses.
However, the finite size effect of $\alpha_a M^2_a$ is important.
The soft mass squared $m_i^2$ converges on  
\[
m^2_{i} \rightarrow \frac{C_{i a}}{\Gamma_{i}} \alpha_a(M_c) M^2_a (M_c),
\]
where we denote 
\[
\frac{C_{i a}}{\Gamma_{i}} = {\cal M}^{-1}_{ij}C_{ja}.
\]
The constant $\Gamma_i$ can be obtained from fixed point values of 
the gauge and Yukawa couplings of the SC sector by fixing a model from 
eq.~(\ref{beta-soft-m}),  
and it is of $O(\gamma_i)$, that is, $\Gamma_i \leq O(1)$.
What is important is that  $\Gamma_i$ is flavor-dependent.
Because anomalous dimensions $\gamma_i$ are flavor-dependent 
to realize hierarchical Yukawa couplings.
Thus, the deference between sfermion masses, 
{\it e.g.} the first and the second families, is obtained by
\begin{equation}
m^2_2(M_c)-m^2_1(M_c) = C_{\tilde f a} \alpha_a(M_c) M^2_a (M_c)
\left({1 \over \Gamma_2} - 
{1 \over \Gamma_1}\right),
\label{dif-12}
\end{equation}
where $C_{i a}$ is denoted by $C_{\tilde f a}$, 
because the quadratic Casimir is common.
Here $\Gamma_2$ would be smaller than $\Gamma_1$ 
to obtain realistic Yukawa matrices.
Naturally, we would have 
$1/\Gamma_2 - 1/\Gamma_1 = O(1/\Gamma_2)$.
Below $M_c$ we have only flavor-blind radiative corrections.
Hence, the mass difference is estimated as eq.~(\ref{dif-12}) 
at any scale below $M_c$.
Actually this difference is suppressed by the one-loop factor 
$\alpha_a$ compared with 
the initial value and that is favorable for FCNC constraints.
However, whether that is indeed suppressed enough for FCNC constraints 
depends on radiative corrections between $M_c$ and 
the weak scale, and an explicit value for $\Gamma_2$.

Before estimating nondegeneracy explicitly for the MSSM, 
we give a comment on the fact we have neglected 
the SC gaugino mass $M'$ and A-parameters $A_\lambda$, 
which correspond to tri-linear couplings among 
the SC sector and the SM sector.
In pure SCFTs without effects of the SM gaugino masses $M_a$, 
all of them decrease exponentially as discussed in section 2.
However, for nonvanishing $\alpha_a M_a$ they converge on 
$M'=O(\alpha_a M_a) $ and $A_\lambda =O(\alpha_a M_a) $.
The RG equations of soft scalar masses 
squared $m_i^2$ include 
the terms of $M'^2$ and $A_\lambda^2$.
These are small compared with $\alpha_a M^2_a$ in eq.~(\ref{rg-m-M}) by 
the loop factor $\alpha_a$.
That justifies our above calculations.

Here we study degeneracy of sfermion masses explicitly for the MSSM.
Sfermion masses in the MSSM are obtained at $M_c$,
\begin{eqnarray}
m_{Qi}^2(M_C) &=& {1 \over \Gamma_{Qi}}
[{16 \over 3}\alpha_3M_3^2 + 3\alpha_2M_2^2 
+{1 \over 15}\alpha_1M_1^2 ](M_C) \\
m_{ui}^2(M_C) &=& {1 \over \Gamma_{ui}}
[{16 \over 3}\alpha_3M_3^2 + {16 \over 15}\alpha_1M_1^2 ](M_C) \\
m_{di}^2(M_C) &=& {1 \over \Gamma_{ui}}
[{16 \over 3}\alpha_3M_3^2 + {4 \over 15}\alpha_1M_1^2 ](M_C) \\
m_{Li}^2(M_C) &=& {1 \over \Gamma_{Li}}
[3\alpha_2M_2^2 +{3 \over 5}\alpha_1M_1^2 ](M_C) \\
m_{ei}^2(M_C) &=& {1 \over \Gamma_{ei}}
[{12 \over 5}\alpha_1M_1^2 ](M_C).
\end{eqnarray}
To be explicit, here we write radiative corrections 
due to gaugino masses 
between $M_c$ and $M_Z$,
\begin{eqnarray}
m_{Qi}^2(M_Z) - m_{Qi}^2(M_C) &=& 
{8 \over 9}[{\alpha^2_3(M_Z) \over \alpha^2_3(M_C) } -1 ] M^2_3(M_c)
+{3 \over 2}[1 - {\alpha^2_2(M_Z) \over \alpha^2_2(M_C) } ] M^2_2(M_c) 
\label{rad-m1} \\
& & +{1 \over 198}[1 - {\alpha^2_1(M_Z) \over \alpha^2_1(M_C) } ] M^2_1(M_c)\\
m_{ui}^2(M_Z) - m_{ui}^2(M_C) &=& 
{8 \over 9}[{\alpha^2_3(M_Z) \over \alpha^2_3(M_C) } -1 ] M^2_3(M_c)
+{8 \over 99}[1 - {\alpha^2_1(M_Z) \over \alpha^2_1(M_C) } ] M^2_1(M_c)\\
m_{di}^2(M_Z) - m_{di}^2(M_C) &=& 
{8 \over 9}[{\alpha^2_3(M_Z) \over \alpha^2_3(M_C) } -1 ] M^2_3(M_c)
+{2 \over 99}[1 - {\alpha^2_1(M_Z) \over \alpha^2_1(M_C) } ] M^2_1(M_c)\\
m_{Li}^2(M_Z) - m_{Li}^2(M_C) &=& 
{3 \over 2}[1 - {\alpha^2_2(M_Z) \over \alpha^2_2(M_C) } ] M^2_2(M_c)
+{1 \over 22}[1 - {\alpha^2_1(M_Z) \over \alpha^2_1(M_C) } ] M^2_1(M_c)\\
m_{ei}^2(M_Z) - m_{ei}^2(M_C) &=& 
{2 \over 11}[1 - {\alpha^2_1(M_Z) \over \alpha^2_1(M_C) } ] M^2_1(M_c).
\label{rad-m5}
\end{eqnarray}
We have assumed to have exactly the MSSM matter content 
below $M_c$.
These radiative corrections are quite large compared with 
the initial values at $M_c$.
Thus, the nondegeneracy,  
$\Delta m^2_{\tilde f} = 
(m^2_{\tilde f2} - m^2_{\tilde f1})/m^2_{\tilde f,av}$,
where $m^2_{\tilde f,av}$ is an average value of them, 
is obtained as 
\begin{equation}
\Delta m^2_{\tilde f} = {C_{\tilde f a} \alpha_a(M_c) M^2_a (M_c) \over 
m^2_{\tilde f}(M_Z) }
\left({1 \over \Gamma_2} - 
{1 \over \Gamma_2}\right).
\end{equation}
To estimate such nondegeneracy, we define 
\begin{equation}
\Delta_{\tilde f} = {C_{\tilde f a} \alpha_a(M_c) M^2_a (M_c) \over 
m^2_{\tilde f}(M_Z) }.
\end{equation}
To be explicit, we use  
\begin{eqnarray}
\Delta_{\tilde Q} &=& 
{(16/3)\alpha^3_3(M_c)+3\alpha^3_2(M_c)+(1/15)\alpha^3_1(M_c) \over -
(8/9)\Delta \alpha^2_3
+(3/2)\Delta \alpha^2_2
+(1/198)\Delta \alpha^2_1}  \\
\Delta_{\tilde u} &=& 
{(16/3)\alpha^3_3(M_c)+(16/15)\alpha^3_1(M_c) \over -
(8/9)\Delta \alpha^2_3
+(8/99)\Delta  \alpha^2_1}  \\
\Delta_{\tilde d} &=& 
{(16/3)\alpha^3_3(M_c)+(4/15)\alpha^3_1(M_c) \over -
(8/9)\Delta \alpha^2_3
+(2/99)\Delta  \alpha^2_1}  \\
\Delta_{\tilde L} &=&
{3\alpha^3_2(M_c)+(3/5)\alpha^3_1(M_c) \over 
(3/2)\Delta \alpha^2_2
+(1/22)\Delta \alpha^2_1}  \\
\Delta_{\tilde e} &=& 
{(12/5)\alpha^3_1(M_c) \over 
(2/11)\Delta \alpha^2_1} = 
 {66 \over 5}{\alpha_1(M_C) \over 1- (\alpha_1(M_Z)/\alpha_1(M_C))^2},
\end{eqnarray}
where $\Delta \alpha^2_i = \alpha^2_i(M_C)-\alpha^2_i(M_Z)$.
Recall that we have assumed gaugino mass unification $M_a=M_{1/2}$ 
at the GUT scale $M_x$.
It should be noted that $\Delta_{\tilde f}$, therefore
$\Delta m^2_{\tilde f}$, may be predicted independently of the
SM gaugino masses.

Fig.~2 shows $\Delta_{\tilde Q}$ and $\Delta_{\tilde d}$ against $M_c$.
We omitted to present $\Delta_{\tilde u}$, because it is almost 
same as $\Delta_{\tilde d}$.
As a result, this mechanism can realize favorable 
degeneracy between squark masses for large $M_c$.
For $\Gamma_i > 0.1$ we could avoid the FCNC problem.
On the other hand, the FCNC problem would be serious 
for smaller values of $\Gamma_i$.

\begin{center}
\input Delta-q.tex

Figure 2: $\Delta_{\tilde Q}$ and $\Delta_{\tilde d}$ against $M_c$.
\end{center}

Similarly Fig.~3 shows $\Delta_{\tilde L}$ and $\Delta_{\tilde e}$ 
against $M_c$.
We have a good degeneracy between left-handed sleptons.
For $\Gamma_i > 0.1$ we could avoid the FCNC problem.
However, for the right-handed slepton the degeneracy is 
not strong compared with squarks and left-handed sleptons.
The reason is that the radiative correction due to the 
bino is not large compared with the others.
In this case, we would face with the FCNC problem for 
$\Gamma_i \sim O(0.1)$.

\begin{center}
\input Delta-l.tex

Figure 3: $\Delta_{\tilde L}$ and $\Delta_{\tilde e}$ against $M_c$.
\end{center}

We have ignored contributions to 
the RG equations due to the $U(1)_Y$ D-term.
However, such contribution would be sizable in particular 
for the right-handed slepton masses.
Therefore we also discuss contributions due to the $U(1)_Y$ D-term.
Including such effects the right-handed slepton mass 
squared $m^2_{\tilde ei}$ at $M_c$  is obtained 
\begin{equation}
m_{ei}^2(M_C) = {1 \over \Gamma_{ei}}
{3 \over 5}\alpha_1[4M_1^2 -S](M_C), 
\end{equation}
where $S={\rm Tr} Ym^2_i$, {\it i.e.},
\begin{equation}
S= m^2_{Hu}-m^2_{Hd} + \sum_i 
(m^2_{\tilde Q i}-2m^2_{\tilde u i}+m^2_{\tilde d i}-
m^2_{\tilde L i}+m^2_{\tilde e i}).
\end{equation}
At $M_c$, the fields which do not couple to the SC sector, 
{\it e.g.} stop and Higgs fields,  
have non-suppressed soft scalar masses, and 
these masses contribute to the initial value of $S(M_c)$, 
which is in general not suppressed and would be of $O(M^2_a)$.
In addition, the radiative corrections including the  $S$-effect
are obtained
\begin{equation}
m_{ei}^2(M_Z) - m_{ei}^2(M_C) =
{2 \over 11}[1 - {\alpha^2_1(M_Z) \over \alpha^2_1(M_C) } ] M^2_1(M_c) 
+ {1 \over 11}[{\alpha_1(M_Z) \over \alpha_1(M_C)} -1] S(M_c).
\end{equation}
Fig.~4 shows $\Delta_{\tilde e}$ including these effects for 
$S(M_c)=0, -M^2_1(M_c)$ and $-10M^2_1(M_c)$.
We have a slight suppression of $\Delta_{\tilde e}$, 
but that is not drastic enough to change its order.
Thus, for $\Gamma_i = O(0.1)$ we would still have 
the serious FCNC problem.
In this case, we may be required to take the degenerate case 
with $\gamma_{e1}=\gamma_{e2}$ and $\Gamma_{e1}= \Gamma_{e2}$.
That would constrain the form of lepton Yukawa matrix.

\begin{center}
\input Delta-es.tex

Figure 4: $\Delta_{\tilde e}$ against $M_c$ with $S=0$, 
$S=-M_1^2$ and $S=-10M_1^2$.
\end{center}

We have assumed the universal gaugino mass $M_a(M_X) = M_{1/2}$.
We can relax the condition.
However, all of the above results on degeneracy of 
sfermion masses are similar.
Because only one of gaugino masses contributes 
almost dominantly to each fermion mass degeneracy, 
that is, $M_3$, $M_2$ and $M_1$ contribute 
to degeneracy of squark masses, left-handed slepton masses 
and right-handed slepton masses, respectively.

We have assumed that the SC region is below $M_X$.
Alternatively, We can take the  possibility that 
the SC region is between $M_X$ and the Planck scale, and 
the Nelson-Strassler mechanism would work in some GUT model.
Such case can be studied similarly and 
we may have a significant change for the 
slepton masses.
Such GUT scenario shall be discussed elsewhere \cite{prog}.

\subsection{Mass spectrum}

Here we show representative mass spectra in the case that 
we have the exactly same matter content below $M_C$ as the MSSM 
and the gaugino masses are unified at $M_X$, $M_a(M_X)=M_{1/2}$.
As we  saw in the previous subsection, sfermion masses can be
quite suppressed at $M_c$ for the fields, 
which couple with the SC sector and whose 
anomalous dimensions are determined definitely.
Namely, we have no-scale type of initial conditions for 
such fields.
Thus magnitudes of sfermion masses of this type are 
calculated only by radiative corrections between $M_c$ 
and the weak scale (\ref{rad-m1})-(\ref{rad-m5}).
Fig.~5  shows ratios of sfermion masses to $M_3$ at the weak scale.
The three solid lines correspond to  
$m_{\tilde Q}/M_3$, $m_{\tilde L}/M_3$ and $m_{\tilde e}/M_3$, 
respectively.
We have taken $S=0$.
Also the two dotted lines show ratios of gauginos 
to $M_3$.
The upper and the lower correspond to $M_2/M_3$ and $M_1/M_3$,
respectively.
Note that the right-handed slepton is lighter than the bino.
In this case the LSP would be slepton and 
the ordinary no-scale type initial condition has the same problem
\cite{lsp,yamaguchi}, 
although we have to take mass eigenvalues and it depends on 
the overall magnitude of soft masses.
However, the $U(1)_Y$ D-term has a sizable effect as 
discussed in subsection 3.2.
Fig.~6 shows $m_{\tilde e}/M_3$ for $S(M_c) = -0.5M^2_3(M_Z)$.
In this case the LSP is the neutralino.

\begin{center}
\input s-spec.tex
\vspace*{2mm}\\
\parbox{160mm}
{Figure 5: Ratios of sfermion masses to $M_3$ at the weak scale.
The three solid lines correspond to  
$m_{\tilde Q}/M_3$, $m_{\tilde L}/M_3$ and $m_{\tilde e}/M_3$, 
respectively. The two dotted (upper and lower) lines correspond to 
ratios of gaugino masses to $M_3$ ($M_2/M_3$ and $M_1/M_3$).
}
\end{center}

\begin{center}
\input spec-es.tex

Figure 6: $m_{\tilde e}/M_3$ for $S=0$ and $-0.5M_3^2$.
\end{center}

The masses of the sfermion which do not couple with the 
SC sector, {\it e.g.} stop masses (and sbottom and 
stau masses for the large $\tan \beta$ scenario), depend on 
their initial conditions.
It is natural to assume their masses are of $O(M_a(M_X))$ or 
$O(M_a(M_C))$.
Note that the ratio $M_3(M_Z)/M_{1/2}$ is less than 3, 
(which is expected in the ordinary MSSM) if 
$M_c$ is lower than $M_X$.
Because the unified gauge coupling becomes large by adding 
extra matter fields.
We have a large mass gap between the stau and the other sleptons
if the stau couple with the SC sector.
On the other hand, whether stop is lighter than the other squarks 
depends on the initial condition.
Anyway, we can predict definitely the mass spectrum 
for the sfermions coupling with the SC sector for 
fixed $M_c$.
Also we could relax the condition with the universal gaugino mass 
$M_a(M_X) = M_{1/2}$.

\section{Analyses of squark masses in explicit models}

\subsection{Models with suppressed soft parameters}
The models based on the SCFT with exponentially suppressed scalar masses
are favorable phenomenologically in avoiding the flavor problems.
In this subsection we consider the perturbatively renormalizable theories 
enjoying this property. 
Indeed we could consider also many varieties by using the SCFTs with higher 
dimensional operators as discussed in section~2. 
However, in that case, we should start with the assumption that  
there exists such IR fixed point, 
because of lack of RG frameworks applicable to non-renormalizable
theories.
Therefore we shall restrict ourselves to the renormalizable theories.
Then the types of models with suppressed scalar masses are 
found to be rather limited as follows.

Suppose a quark (lepton) $q$ couples to the SCFT through Yukawa 
interaction $qQP$.
Then we seek for the models in which $\gamma_Q+\gamma_P$ is uniquely 
determined 
by the fixed point conditions in terms of the anomalous dimensions.
In this case the squark (slepton) mass decreases exponentially as is 
shown in section~2.
Now the interactions are limited to the Yukawa type in renormalizable theories.
Here let us also assume that there is no Yukawa terms composed only of
non-singlet fields under $G_{SC}$.
Then $\gamma_Q+\gamma_P$ must be determined by the condition for vanishing
gauge beta function.
This means that the gauge beta function should depend only of 
$\gamma_Q + \gamma_P$.
On the other hand the quadratic Casimirs of $Q$ and $P$ must be equal, 
since $QP$ forms a $G_{SC}$-singlet.
Therefore the dimensions of $G_{SM}$ representations of $Q$ and $P$ are also
necessarily the same.

Taking into account the fact that $q$ carries $G_{SM}$  charges, 
possible types of the models
seem to be rather limited.
We shall enumerate a few simple examples below.
\begin{enumerate}
\item Chiral $SU(5)$ model\\
The SC-gauge group $G_{SC}$ is $SU(N_c)$ and the SM-gauge group $G_{SM}$ is $SU(5)$.
We introduce the following chiral fields assigned the representations under 
$(G_{SC}, G_{SM})$;
\be
Q : (N_c, \bar{5}),~~~ P : (\bar{N}_c, \bar{5}),~~~ q : (1, 10).
\ee
The superpotential is given by $W = \lambda q Q P$.
Then the IR fixed point is found to exist for $N_c=2, 3$.

In this class of models the SC-gauge non-singlet fields $Q$ and $P$
belong to the same dimensional representations of the SM-gauge group
and, therefore, their anomalous dimensions are equal.
\footnote{The $SU(3)^3$ model in ref.~\cite{ns} belongs to this class.}
All scalar masses, $m^2_Q, m^2_P$ and $m^2_q$ converges to 0 irrespective of
initial values.

\item L-R symmetric $SU(3)$ model\\
Suppose $G_{SC}=SU(N_c)$ and $G_{SM}=SU(3)$ and introduce
\bea
& &Q : (N_c, 3),~~~ \bar{Q} : (\bar{N}_c, \bar{3}), ~~~
P : (\bar{N}_c, 3), ~~~\bar{P} : (N_c, \bar{3}), \nn \\
& &q_L : (1, 3),~~~ q_R : (1, \bar{3}).
\eea
Also the superpotential is defined as
$W=\lambda (q_L Q \bar{P} + q_R \bar{Q} P)$.
The IR fixed point is found to exist for $N_c=3$.

The anomalous dimensions of $Q$ and $\bar{Q}$, also $P$ and $\bar{P}$ are
the same by the left-right symmetry.
Therefore $\gamma_Q + \gamma_P$ is fixed by the fixed point equation
given by eq.~(\ref{eq:fpcond}).
In such cases, however, we need to assume 
$m^2_{Q}=m^2_{\bar{Q}}, m^2_{P}=m^2_{\bar{P}}$ for exponential
suppression of the scalar masses $m_{q_L}$ and $m_{q_R}$.
Note that $m^2_{Q}$ or $m^2_{P}$ is not reduced to 0, though the sum 
of them decreases exponentially.

\end{enumerate}

For these types of models we cannot introduce two quarks with distinct anomalous 
dimensions in a single SC-gauge sector. 
\footnote{
The hierarchy of Yukawa couplings can be generated by assuming different decoupling
scales $M_c$ instead of the anomalous dimensions. In such cases we may make
several quarks couple to a common SC-sector.
}
In other words we need to assume different
SC-gauge theory for every quark or lepton to be given large anomalous dimension.
There may be some exceptional cases that the Yukawa interactions composed only of 
the SC-gauge non-singlet fields are also allowed.
In this paper we are not going to explore such possibilities here.
Hereafter we discuss the IR behavior of the soft scalar masses by considering
the models akin to the above examples.

\subsection{Sfermion mass convergence in $SU(3)_{SC}\times SU(3)_C$ model}
In  subsection 3.2 we have evaluated the flavor dependence of
squark masses.
In this discussion we have assumed that soft scalar masses in the SC-sector
converge sufficiently.
However, the range of scale where the theory is regarded as a SCFT
must be finite, otherwise the Yukawa couplings are suppressed out too much.
Therefore soft scalar masses would not converge completely either
at the decoupling scale $M_c$.
Degree of the convergence is related to the suppression for the
Yukawa couplings.
First let us estimate roughly how much the squark masses converge.

Suppose that the theory is regarded as a SCFT at the scale of
$M_c < \mu < \Lambda_c$.
In this region the soft scalar masses are subject to eq.~(\ref{rg-m-M}) 
ignoring the gaugino mass and A-parameter in the SC sector.
Then the speed of convergence is given by the smallest eigenvalue $\xi$
of the matrix ${\cal M}$.
This eigenvalue is found to be the same order of the anomalous dimensions
of $\phi_i$.
Let us define the deviation of the squark mass from the convergent value
by $\delta m^2_i = m^2_i- (C_i/\Gamma_i)\alpha_3 M^2_3$.
Then the deviation at $M_c$, which is estimated roughly as 
\be
\delta m^2_i(M_c)
= e^{-\xi_i \ln(\Lambda_c/M_c)}
\delta m^2_i(\Lambda_c),
\ee
has to be much less than $\alpha_3 M^2_3$
in order  that the formula for the squark masses given in the previous section
are valid.
Also if $\delta m^2_i$ is found to be much larger than $\alpha_3 M^2_3$,
the squark masses may not be degenerate enough so as to avoid the flavor 
problem.

The ratio of the Yukawa couplings is determined by the anomalous dimension
of the quarks. 
By noticing that the eigenvalue $\xi_i$ is found to be the same order as
the anomalous dimension, we evaluate $\delta m^2_i$ also as
\be
\delta m^2_i(M_c) 
\sim \frac{y_{ii}(M_c)}{y_{ii}(\Lambda_c)} m^2_i(\Lambda_c) \\
\sim \frac{m_{q_i}}{m_{q_3}}  m^2_i(\Lambda_c),
\ee
where $m_{q_i}$ denotes the quark mass of $i$-th generation.
Especially the deviation $\delta m^2_i$ for the second generation
should be suppressed by a factor similar to $m_s/m_b \sim O(10^{-2})$.
Therefore there may remain a large uncertainty in the squark mass due to this 
deviation at $M_c$ for the second generation.
If the squark mass is the same order as the SM-gaugino mass at $\Lambda_c$,
this uncertainty is supposed to be much larger than the convergent value 
evaluated in the previous section.
Therefore the SM-gaugino mass is required to be fairly larger than the
squark masses at $\Lambda_c$.

In practice the above argument is rather bold. 
In the followings we shall demonstrate the RG flows for the squark masses and
their converging behavior explicitly in a concrete model and examine
the convergence.
Suppose both of the SC-gauge and SM-gauge groups are $SU(3)$
and introduce the following chiral fields,
\bea
& & Q=(3, \bar{3}),~~~\bar{Q}=(\bar{3}, 3), ~~~
P=(3, 3),~~~\bar{P}=(\bar{3}, \bar{3}), \nn \\
& & q_i=(1, 3),~~~\bar{q_i}=(1, \bar{3}) ~~~(i=1, 2, 3),~~~H=(1, 1).
\eea
The superpotential is defined by
\be
W=\lambda (q_1 \bar{Q}P + \bar{q}_1\bar{P}Q) + y_i \bar{q_i}q_i H.
\ee
Here we have simplified the Yukawa couplings to the diagonal ones.
In this toy model only the Yukawa coupling of the first generation,
$y_1$ is suppressed.
Also we assume $m^2_{Q}=m^2_{\bar{Q}}$ and $m^2_{P}=m^2_{\bar{P}}$.

Below we analyze the RG flows of the various couplings numerically
by substituting the anomalous dimensions
in the exact beta functions with those evaluated in 1-loop perturbation.
The anomalous dimensions are given by
\bea
\gamma_Q &=& 
\gamma_P = -\frac{8}{3}\alpha' + 2\alpha_{\lambda} - \frac{8}{3}\alpha, \\
\gamma_{q_1} &=& 6\alpha_{\lambda} - \frac{8}{3}\alpha + \alpha_{y_1}, ~~~
\gamma_{q_i} = - \frac{8}{3}\alpha + \alpha_{y_i}~~~(i=2, 3), \\
\gamma_{H} &=& 3(\alpha_{y_1}+\alpha_{y_2}+\alpha_{y_3}),
\eea
where $\alpha' = g'^2/8 \pi^2$, $\alpha = g^2/8 \pi^2$, 
$\alpha_{\lambda} = |\lambda|^2/8 \pi^2$ and $\alpha_{y_i}=|y_i|^2/8 \pi^2$.
It is straightforward to derive the beta functions for all couplings by
using formula shown in section~2.
Here let us write down only the beta functions for soft parameters 
in the SC-sector;
\bea
\mu \frac{dM'}{d\mu} &=&
-\frac{3 \alpha'(2-3\alpha')}{(1-3\alpha')^2} [1+ 2\gamma_Q]M'
-\frac{6 \alpha'^2}{1-3\alpha'} \gamma^{(1)}_Q,\\
\mu \frac{dA_{\lambda}}{d\mu} &=&
-(2\gamma^{(1)}_Q + \gamma^{(1)}_{q_1}), \\
\mu \frac{d m^2_Q}{d \mu} &=&
\mu \frac{d m^2_P}{d \mu} = \gamma^{(2)}_Q, \\
\mu \frac{d m^2_1}{d \mu} &=& \gamma^{(2)}_{q_1},
\eea
where $\gamma^{(1)}$ and $\gamma^{(2)}$ are obtained by the superfield
extension discussed in section~2. By neglecting terms of $O(\alpha^2)$ 
or of $O(\alpha_{y1})$ as negligible amounts, they are given by
\bea
\gamma^{(1)}_Q &=&
-\frac{8}{3} \alpha'M' -2\alpha_{\lambda}A_{\lambda} -\frac{8}{3}\alpha M,\\
\gamma^{(1)}_{q_1} &=&
-6\alpha_{\lambda}A_{\lambda} - \frac{8}{3}\alpha M,\\
\gamma^{(2)}_Q &=&
-\frac{8}{3}\alpha'(2|M'|^2 + \Delta_{g'}) 
+ 2 \alpha_{\lambda}(|A_{\lambda}|^2 + m_Q^2 + m_P^2 + m_1^2)
-\frac{16}{3}\alpha |M|^2, \\
\gamma^{(2)}_{q_1} &=&
6\alpha_{\lambda}(|A_{\lambda}|^2 + m_Q^2 + m_P^2 + m_1^2)
-\frac{16}{3} \alpha |M|^2,
\eea
where $\Delta_{g'} = 3\alpha'(|M'|^2 - m_Q^2 - m_P^2)$ as defined by 
eq.~(\ref{Delta_g}).

The fixed points are found at 
A: $(\alpha'_*, \alpha_{\lambda *}) = (5/16, 1/6)$ and
B: $(\alpha'_*, \alpha_{\lambda *}) = (3/16, 0)$.
The point A is the IR attractive fixed point and 
the anomalous dimensions there are found to be
\[
\gamma_{Q *} = \gamma_{P *} = \frac{1}{2},~~~\gamma_{q_1 *}= -1.
\]
In the region that $M'$ and $A_{\lambda}$ are suppressed to negligible
amounts, the RG evolution of the sfermion masses are given by
\be
\mu \frac{d }{d \mu}
\left(
\begin{array}{c}
m^2_Q + m^2_P \\
m^2_1
\end{array}
\right)
=
\left(
\begin{array}{cc}
16 \alpha'^{2}_*+ 4 \alpha_{\lambda *} & 4 \alpha_{\lambda *} \\
6 \alpha_{\lambda *} & 6 \alpha_{\lambda *}
\end{array}
\right)
\left(
\begin{array}{c}
m^2_Q + m^2_P \\
m^2_1
\end{array}
\right)
-\frac{16}{3}\alpha |M|^2
\left(
\begin{array}{c}
2 \\
1
\end{array}
\right).
\ee
Note that $m^2_Q+m^2_P$ but not each of $m^2_Q$ and $m^2_P$ converges to
$O(\alpha |M|^2)$ in this model.
When $\alpha |M|^2$ is treated as a constant, the eigenvalues of this 
coupled equation are found to be $(2.64, 0.59)$.
Indeed the smaller one $\xi=0.59$ is close to the anomalous dimension
$1/2$.
Therefore degrees of suppression for the Yukawa coupling
and the scalar masses are almost same in this model.
It is also expected that the scalar masses converge 
as $m^2_1 \rightarrow 0.78 \alpha M^2$, 
$m^2_Q+m^2_P \rightarrow 4.55 \alpha M^2$. 

Now we present the results obtained by numerical analyses
of the RG equations.
In Fig.~7 the aspect of suppression for $(y_1, M', A_{\lambda})$
are shown with respect to $t=\log_{10}(\mu/\Lambda_c)$.
Here we set $\alpha'$ and $\alpha_{\lambda}$ on the IR fixed point.
The initial values for other couplings are chosen as follows:
$M'=A_{\lambda}=1.0, M=5.0, \alpha=1/(48\pi), 
\alpha_{y1}=1/(8\pi^2)$.
The value of $\alpha$ refers to the GUT gauge coupling.
It is seen that the Yukawa coupling is smoothly suppressed.
If we suppose $M_c$ to be the scale that the Yukawa coupling is 
suppressed by $10^{-2}$, then $t_c=\log_{10}(M_c/\Lambda_c)$
is found to be $-2.01$. 

\setcounter{figure}{6}
\begin{figure}[htb]
\begin{center}
\epsfxsize=0.6\textwidth
\leavevmode
\epsffile{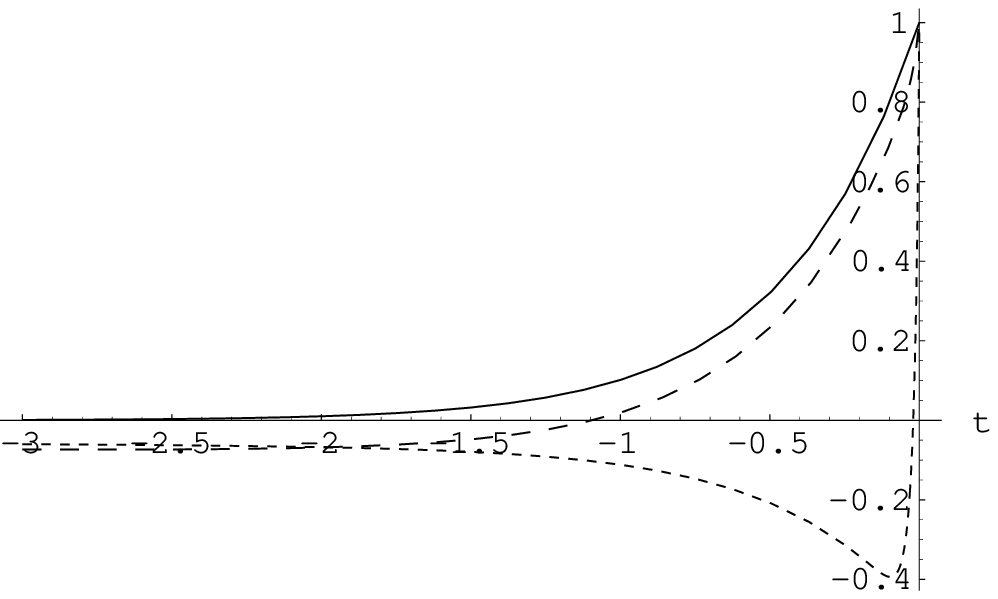}
\caption{The running couplings 
$(y_1, M', A_{\lambda})$ are shown 
in ratio to their initial values by
a solid, a dashed and a long-dashed line respectively.
$t=\log_{10}(\mu/\Lambda_c)$.
}
\end{center}
\end{figure}

Next we examine the RG flows of the sfermion masses by varying the
initial values and observe the converging behavior.
Fig.~8 shows the RG flows obtained by varying the initial value for 
$m^2_1$ between $[0.0, 2.0]$ with setting $m^2_Q=m^2_P=1.0$.
It is seen that the sfermion masses converges to the values of
$O(\alpha |M|^2)$, though the coefficients are slightly shifted
from the above naive estimation; $m^2_1 \rightarrow 0.3 \alpha M^2$.
It is found also that the range of $m^2_1$ shrinks to about 5\% 
of initial one at $M_c$.
Actually we obtain the almost same results for any setting for the
initial couplings.
For generic initial sfermion masses of the  same order of the 
SM-gaugino mass given at $\Lambda_c$, then the deviation
$\delta m^2_i$ is found to remain about 10 times larger than the 
converging value at $M_c$.
Thus we conclude that the strongly degenerate squark mass spectrum
evaluated in section~3 is indeed achieved irrespective of the initial sfermion
masses, if the SM-gaugino mass is fairly larger than them.

\begin{figure}[htb]
\begin{center}
\epsfxsize=0.6\textwidth
\leavevmode
\epsffile{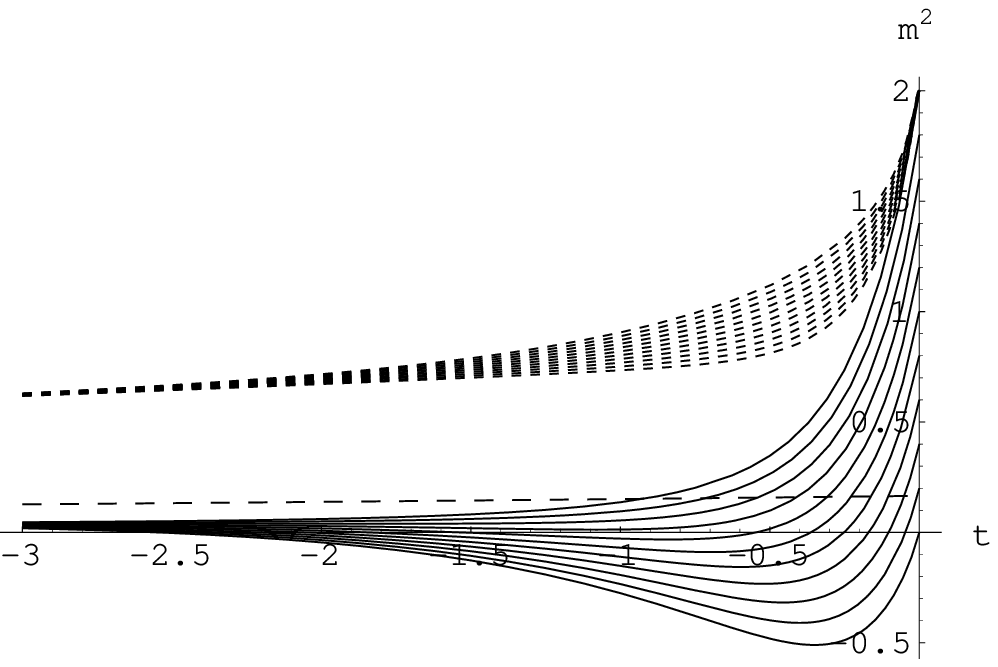}
\caption{RG flows for $(m^2_1, m^2_Q+m^2_P)$ shown by
solid and dashed lines respectively.
The long dashed line gives $\alpha M^2$.
}
\end{center}
\end{figure}

In practice the theories must become SCFTs at a certain scale in
order to generate finite ratio among Yukawa couplings.
Therefore we have also performed the similar observations by assuming the
initial vales of $\alpha'$ and $\alpha_{\lambda}$ off the fixed point
at higher energy scale $\Lambda_0 > \Lambda_c$.
However the results obtained on the convergence for the sfermion masses
are not significantly changed.

\section{Generation of $\mu$-term by a singlet}
So far, we have discussed the cases that
the large anomalous dimensions for quarks and leptons are 
determined definitely. 
It has been seen that the corresponding sfermion masses
are exponentially suppressed and converge to non-vanishing
values due to effects of SM-gaugino mass.
It will be shown that this converging value for (mass)$^2$ can be 
negative for a singlet field coupled with the SC sector.
The order of the tachyonic (mass)$^2$ is fixed to $O(\alpha M^2)$,
namely the weak scale, irrespectively of the bare scalar mass.
On the other hand the weak scale mass term ($\mu$-term) in the 
supersymmetric SM has no theoretical grounds and poses the so-called  
$\mu$-problem \cite{mu-prob}.
It has been discussed sometimes that a singlet can explain
the $\mu$-term by developing its vacuum expectation value (VEV) of the
weak scale \cite{singlet,extraU1}.
In this section we propose another solution for the $\mu$-problem
by considering a singlet field coupled with the SC sector.

Suppose that a singlet field $S$ coupled to the SC-sector
through the superpotential
\footnote{
Here we assume the bare $\mu$-term is absent in the superpotential.
Indeed the $\mu$-term may be prohibited by imposing a discrete symmetry. 
However the discrete symmetries lead to a cosmological problem
by forming domain walls in general.
Alternatively we may introduce an extra $U(1)$ gauge symmetry to
forbid the bare $\mu$-term.
}
\be
W=S Q \bar{Q} + S^3.
\ee
Here we also assume that $(Q, \bar{Q})$ carry SM-gauge charges.
By assuming $m^2_Q=m^2_{\bar{Q}}$, the form of the RG equations 
for the scalar masses are given generally as
\bea
\mu \frac{d m^2_Q}{d \mu} &=& (a + 2b)m^2_Q + b m^2_S 
- C_Q \alpha |M|^2, \nn \\
\mu \frac{d m^2_S}{d \mu} &=& 2c m^2_Q + c m^2_S,
\eea
where the gaugino mass and A-parameter are ignored again.
The coefficients $a, b, c$ and $C_Q$ are positive 
and determined by group-theoretical factors.

If $Q$ and $\bar{Q}$ are SM-gauge singlets, or $C_Q=0$, the
scalar masses $m^2_Q$ and $m^2_S$ are reduced to 0 exponentially.
However the correction by the SM-sector gaugino, $\alpha |M|^2$,
makes the scalar masses converge to non-vanishing values.
We can treat the gaugino mass as well as the gauge coupling in the
SM-sector as constants, since their evolution is slow enough.
Then the scalar masses converge to
\bea
m^2_Q &\rightarrow& \frac{C_Q}{a}\alpha |M|^2,  \nn \\
m^2_S &\rightarrow&  -\frac{2C_Q}{a}\alpha |M|^2.
\eea
Here we should note that the singlet $S$ becomes tachyonic
irrespective of the initial values of the scalar masses.
The singlet mass remains to be tachyonic and also appears in
$O(\alpha M^2(M_c))$ at the weak scale.
Thus the singlet field $S$ necessarily acquires a VEV
of this order through coupling to the SC-sector.
If the Yukawa coupling $SH_u H_d$ exists, 
the $\mu$-term is generated as the order of the weak scale 
automatically.

The singlet field generating the $\mu$-term of the weak
scale can be incorporated with the SC-sector inducing 
Yukawa suppression.
Here let us demonstrate this by introducing a singlet to the
$SU(3)_{SC}\times SU(3)_C$ model analyzed in 
section~4.
We extend the superpotential of the SCFT as
\be
W= \lambda(q_1\bar{Q}P + \bar{q}_1 \bar{P} Q) + y_{ij}\bar{q}_i q_j H
+ \lambda' S\bar{Q}Q + \lambda''S \bar{P}P  + S H^2 + S^3.
\ee
At the IR fixed point the anomalous dimensions are fixed
to be $\gamma_Q=\gamma_P=-1/2, \gamma_{q_1}=\gamma_S=1$.
It is seen that $\lambda'=\lambda''$ at the fixed point
from $\gamma_Q=\gamma_P$.

Below we examine the RG equations by applying the anomalous 
dimensions obtained by 1-loop perturbation:
\bea
\gamma_Q &=&
-\frac{8}{3} \alpha' + 2 \alpha_{\lambda} + \alpha_{\lambda'} -\frac{8}{3}\alpha,\\
\gamma_P &=&
-\frac{8}{3} \alpha' + 2 \alpha_{\lambda} + \alpha_{\lambda''} -\frac{8}{3}\alpha,\\
\gamma_{q_1} &=&
6 \alpha_{\lambda} -\frac{8}{3}\alpha,\\
\gamma_S &=&
3 \alpha_{\lambda'} + 3 \alpha_{\lambda''}.
\eea
The IR fixed point couplings are found at
$\alpha'_{*}=3/16, \alpha_{\lambda *}=\alpha_{\lambda' *}=1/6$.
By ignoring the gaugino mass and A-parameter of the SC-sector again,
the RG equations for the scalar masses are given by
\be
\mu \frac{d}{d \mu} 
\left(
\begin{array}{c}
m^2_Q +m^2_P \\ m^2_{q_1} \\ m^2_S
\end{array}
\right)
=
\left(
\begin{array}{ccc}
13/4 & 2/3 & 1/3 \\
1 & 1 & 0 \\
1 & 0 & 1
\end{array}
\right)
\left(
\begin{array}{c}
m^2_Q +m^2_P \\ m^2_{q_1} \\ m^2_S
\end{array}
\right)
- \frac{16}{3}\alpha M^2
\left(
\begin{array}{c}
2 \\ 1 \\ 0
\end{array}
\right),
\ee
where the fixed point couplings are used.
{}From this equation it is found that the scalar masses converge
as
\be
\left(
\begin{array}{c}
m^2_Q +m^2_P \\ m^2_{q_1} \\ m^2_S
\end{array}
\right)
\rightarrow
\frac{16}{3}\alpha M^2 
\left(
\begin{array}{c}
16/27 \\ 11/27 \\ -16/27
\end{array}
\right).
\ee
Thus it is seen that the singlet becomes tachyonic indeed.

As another possibility for a singlet to develop the weak scale
VEV, we may consider the SCFTs whose anomalous dimensions are
not uniquely determined by the fixed point conditions 
(\ref{eq:fpcond}).
In such cases the sfermion masses converge to certain values
of the same order as the initial masses.
Therefore the singlet field can be driven to be tachyonic 
by the Yukawa coupling to the SC sector.
However, the converging values depend on the initial
conditions and, hence, it is not automatic for the singlet 
to become tachyonic contrary to the above case.

\section{Conclusions and discussions}

We have studied soft SUSY breaking parameters in 
the Nelson-Strassler type of models: SSMs coupled with 
SCFTs.
We have clarified the condition to derive 
exponentially suppression of sfermion masses 
within the framework of pure SCFTs, 
that is, we have suppressed sfermion masses 
for the fields whose anomalous dimensions are 
determined definitely.

In a realistic case with non-vanishing gauge couplings 
of the SM sector, however, 
the terms $\alpha_a M^2_a$ in RGEs of 
sfermion masses play an important role to realize degenerate 
sfermion masses.
The sfermion masses converge on $O(\alpha_a M^2_a)$ and 
these are flavor-dependent unless $\gamma_i = \gamma_j$.
We have shown explicitly how much degeneracy we have 
between sfermion masses in the MSSM.
For squarks we can have suppression strong enough 
to avoid the FCNC problem.
On the other hand, for sleptons 
such suppression is weak.
For squarks this mechanism is attractive even if 
we could not obtain sufficiently realistic Yukawa matrices only by 
the Nelson-Strassler mechanism, 
that is, it might be useful to introduce a SC sector in order 
only to suppress initial non-degeneracy between 
squark masses.

We have assumed that the SC region is below $M_X$.
It is also possible that the SC region is above $M_X$ and 
the Nelson-Strassler mechanism would work within the GUT framework.
Such case can be studied similarly and 
we would have a significant change for the 
slepton masses.
Such GUT scenario shall be discussed elsewhere \cite{prog}.

Also we have discussed the possibility for generating the 
$\mu$-term.
We can have naturally the singlet fields 
which have tachyonic masses of $O(M_Z)$ and 
whose VEVs generate the supersymmetric mass term 
of the Higgs fields.
It might be possible that a similar mechanism 
generates mass terms of the SC matter fields, 
so that they would decouple the SC sector from the 
SM sector.
This decoupling scale of the SC sector is of $O(M_Z)$, 
$M_c=O(M_Z)$.
That has the problem of the blow-up of $g_a$ 
as discussed in subsection 3.1, if those are 
charged under $G_{SM}$.

Moreover, an application to the neutrino sector 
is interesting.
Since the righ-handed neutrino is $G_{SM}$-singlet, 
we have less limitation for model building.
Such application would be studied elsewhere.

We have studied mainly the degenerate solution for the 
FCNC problem.
Finally we give a comment on the decoupling solution.
It has been shown that the sfermion masses exponentially damp 
in the case that their anomalous dimensions are determined definitely.
Otherwise, squark/slepton masses are of the same order as 
initial values.
Suppose that soft SUSY breaking terms appear only in the SC sector 
including squarks/sleptons coupled with this sector,
while the SM sector has no SUSY breaking terms, that is, 
$M_a=0$ for the gaugino masses of the SM sector and 
$m^2_i=0$ for the stop as well as for the sbottom and stau 
for the large $\tan \beta$ scenario.
In this case, the gaugino and the stop field of the SM sector gain 
masses due to higher loop effects from the SC sector.
Thus, those masses are suppressed by loop factors compared with 
the squark masses of the first and the second families.
That is one of possibilities to realize the decoupling solution.
However, note that although 
squark masses of the first and the second families appear
in the same order as initial values in general, 
the sign of (mass)$^2$  as well as the values are totally
dependent on initial sfermion masses in the SC sector.
We must choose the initial conditions to avoid tachyonic 
sfermion masses.

\section*{Acknowledgements}

The authors would like to thank J.~Kubo and K.~Yoshioka 
for useful discussions.

\end{document}